# Reconfigurable non-Abelian geometric phase in hybrid integrated photonics

Youlve Chen[1]†, Jiaxin Zhang[2]†, Jinlong Xiang[1], An He[1], Junying Li[2*],Yikai Su[1], and Xuhan Guo[1*]

[1]State Key Laboratory of Photonics and Communications, School of Information and Electronic Engineering, Shanghai Jiao Tong University; Shanghai, 200240, China.
[2]Hangzhou Institute for Advanced Study, University of Chinese Academy of Sciences; Hangzhou, 310024, China.

†These authors contributed equally to this work.

**Abstract**

The non-Abelian geometric phase possesses the capability of enabling robust and fault-resilient unitary transformations, making it a cornerstone of holonomic quantum computation. This 'all-geometric' approach has successfully advanced the manipulation of electrons in condensed matter physics and has sparked growing interest in its implementation within photonics — an area that has traditionally relied on sensitive dynamic phases. However, a major limitation of the topologically protected and inherently robust geometric phase is its lack of reconfigurability. In contrast, mainstream optical computing schemes demand high reconfigurability to compensate for fabrication errors and to support diverse computational tasks. Here, we demonstrate a reconfigurable non-Abelian geometric phase based on the non-volatile phase-change material $Sb_2Se_3$. By switching between its crystalline and amorphous states, the number of degenerate subspaces can be actively adjusted. Thus, multilevel second-order matrices and reconfigurable third-order matrices with 3-bit control is realized. For larger reconfigurable rotation angles, tunable braiding operations are also demonstrated. Furthermore, high-dimensional reconfigurable braiding shows promising potential for applications in optical switching. Our results pave the way for the all-geometric-phase-based approach in optical computing.

**Introduction**

The non-Abelian geometric phase is well known for its inherent robustness in realizing unitary transformations. Building upon the foundational work on the Berry phase, Frank Wilczek pointed out that a matrix-valued geometric phase can emerge from the adiabatic evolution of degenerate quantum states. Unlike the dynamic phase, which depends on the system's energy and evolution time, the geometric phase is determined purely by the path traced in the Hilbert space. This unique property underpins the 'all-geometric' approach, which serves as the foundation for fault-tolerant holonomic quantum computing in condensed matter systems. Recently, classical wave systems—such as light and sound—with their rich and versatile degrees of freedom, have emerged as promising platforms for exploring non-Abelian geometric phases. In this context, the 'all-geometric-phase' approach offers advantages including broad operational bandwidth and enhanced fabrication tolerance in optical computing applications. However, the very robustness that makes geometric phases appealing also introduces a fundamental challenge: it inherently limits their reconfigurability. As a result, achieving real-time tunability remains elusive—a key limitation when compared to the requirements of modern programmable photonic circuits for in-situ training and general-purpose optical computing. Thus, the next major goal is to develop strategies that enable highly reconfigurable geometric operations, while maintaining their intrinsic robustness for arbitrary unitary transformations.

Here, we demonstrate a reconfigurable non-Abelian geometric phase by directly manipulating the degenerate subspace, based on a phase-change material (PCM) waveguide platform. This approach simultaneously achieves both robustness and flexible tunability. PCMs have recently emerged as promising candidates for programmable photonic circuits and memory devices. The optical properties of PCMs—such as refractive index and extinction ratio—exhibit significant contrast between their crystalline and amorphous phases. These nonvolatile phase transitions can be triggered optically or electrically, enabling persistent state retention without continuous power input. With advantages including low power consumption, high refractive index contrast, and foundry-compatible, PCMs offer a versatile and scalable platform for integrated photonic applications.

In this work, we demonstrate reconfigurable non-Abelian holonomy based on three-layer hybrid photonic platform incorporating PCM waveguides, specifically utilizing $Sb_2Se_3$. To be specific, some waveguide is made of phase change material, by switching the phase state of the $Sb_2Se_3$ waveguide between its crystalline and amorphous forms, its refractive index undergoes a significant and reversible change, further change the number of degenerate subspace of the Hamiltonian. The range of tunability can be controlled through designing the coupling between silicon waveguide and $Sb_2Se_3$ waveguide. As a result, we achieve reconfigurable non-Abelian geometric operations, including realizations of ordinary SO(2) transformations and two-mode braiding. Furthermore, multi-level tunability is available through the cascading of a series of basic SO(2) transformation components. For-higher dimensional matrices, 8-level (3 bits) reconfigurable SO(3) transformation is synthesized by Givens rotation. Moreover, the proposed architecture supports high-dimensional non-Abelian braiding, which can be harnessed to construct scalable optical switch meshes with potential applications in programmable photonic circuits and quantum information processing. Our work provides a promising pathway toward practical implementations of non-Abelian geometric phases in integrated photonics, opening new avenues for advanced optical computing and reconfigurable quantum technologies.

**Result**

We implement the reconfigurable non-Abelian holonomy in three-layer hybrid integrated photonics platform incorporating PCM. As shown in Fig. 1, all waveguides share an identical cross-sectional geometry. The top waveguide is composed of the phase-change material $Sb_2Se_3$, while the remaining waveguides are made of amorphous silicon. The Hamiltonian of this M-pod structure can be written as:

$$\hat{H}(z) = \begin{bmatrix} \beta & \kappa_{AX}(z) & \kappa_{BX}(z) & \kappa_{CX}(z) & \kappa_{DX}(z) \\ \kappa_{AX}(z) & \beta & 0 & 0 & 0 \\ \kappa_{BX}(z) & 0 & \beta & 0 & 0 \\ \kappa_{CX}(z) & 0 & 0 & \beta & 0 \\ \kappa_{CX}(z) & 0 & 0 & 0 & \beta_P \end{bmatrix} \quad (1)$$

The number of degenerate states supported by this Hamiltonian can be actively controlled by switching the phase state of the $Sb_2Se_3$ waveguide. When $Sb_2Se_3$ is in the amorphous phase (Phase 2 in Fig. 1A), three degenerate states emerge due to the refractive index similarity between amorphous $Sb_2Se_3$ and amorphous silicon (resulting in $\beta = \beta_P$). In contrast, when $Sb_2Se_3$ transitions to its crystalline phase (Phase 1 in Fig. 1A), only two degenerate states remain, as a result of the significant refractive index mismatch between crystalline $Sb_2Se_3$ and amorphous silicon (resulting in $\beta \neq \beta_P$). These two configurations correspond to distinct Hamiltonians that give rise to two different non-Abelian geometric phases through the associated holonomies. In Phase 1, the system exhibits behavior analogous to the previously reported SO(2) holonomy: three couplings is isomorphic to a 2-sphere, the solid angle enclosed by the holonomic path determine the unitary transformation $e^{i\theta\sigma_y}$ between $|D_1\rangle$ and $|D_2\rangle$.

In Phase 2, where three degenerate states ($|D_1\rangle$, $|D_2\rangle$ and $|D_3\rangle$) are preserved, the BWZ connection components $A_{3i} = A_{i3} = 0$ ($i$=1,2,3), since $\kappa_{AX}$ and $\kappa_{DX}$ remains invariant during the evolution, as detailed in the Supplementary Information. Thus, this SO(3) transformation is also equivalent to the rotation generated by the Pauli matrix $\sigma_y$ between $|D_1\rangle$ and $|D_2\rangle$. The accumulated geometric phase $\alpha$ differs from the phase $\theta$ observed in Phase 1 because it involves the participation of $\kappa_{DX}$. Thus, the value of $\kappa_{DX}$ determines the range of this two-level reconfigurable geometric phase.

To achieve a broader tunability, we further design a reconfigurable braiding protocol—where the geometric phase can be modulated from π/2 to 0. As shown in Fig. 1B, the structure consists of three STIRAP processes. When waveguide D is in Phase 1 (refractive index mismatched), standard two-mode braiding occurs. However, when waveguide D transitions to Phase 2 (with matched refractive indices across all waveguides), since the configuration sets $\kappa_{DX} \gg \kappa_{AX}$, $\kappa_{BX}$ and $\kappa_{CX}$, the geometric phase collapses to zero.

Figure 2A presents the experimental results demonstrating a two-level reconfigurable SO(2) holonomy. By switching the phase state of the PCM waveguide, the holonomic path traverses distinct degenerate subspaces, enabling the realization of different SO(2) holonomies. Fig. 2B shows the experimental implementation of reconfigurable two-mode braiding. The phase state of the PCM waveguide determines whether the transformation corresponds to a 'swap' or 'non-swap' operation in the resulting matrix. To demonstrate multi-level discrete reconfigurability, Fig. 2C illustrates the cascading of two independently tunable SO(2) building blocks. By adjusting the phase states of each individual unit, a composite transformation with multi-level tunability is achieved.

To show the scalability, high-dimensional reconfigurable mesh is demonstrated in Fig. 3. A 3-bit (8 level) SO(3) transformation is synthesized in the framework of Givens rotations. The phase state of the PCM waveguide is controlled using laser direct writing. The crystalline and amorphous states can be distinguished by the Raman spectroscopy of their distinct surface textures, as shown in the microscope images provided in the Supplementary Information. The mathematical models, experimentally measured matrix elements magnitude, as well as the fidelity are also shown.

**Discussion**

High reconfigurability is a key enabler for on-chip photonic computing, and programmable photonic circuits have emerged as the mainstream platform for in-situ training and multi-task optical computation, such as Mach-Zehnder interferometers, micro-ring arrays, and intensity modulation architectures. However, the lack of reconfigurability remains a major obstacle in realizing fully geometric ('all-geometric') approaches for photonic computing. In this work, we utilize PCM to demonstrate reconfigurable non-Abelian geometric phase on three-layer hybrid photonic platform. Multi-level, large tunability range SO(m) matrix-valued geometric phase is realized. Furthermore, the reconfigurable two-mode braiding enables flexible switching between 'bar' and 'cross' states, showing promise for large-scale optical switch networks. Our work paves the way for practical applications of non-Abelian geometric phases in photonic computing and information processing.

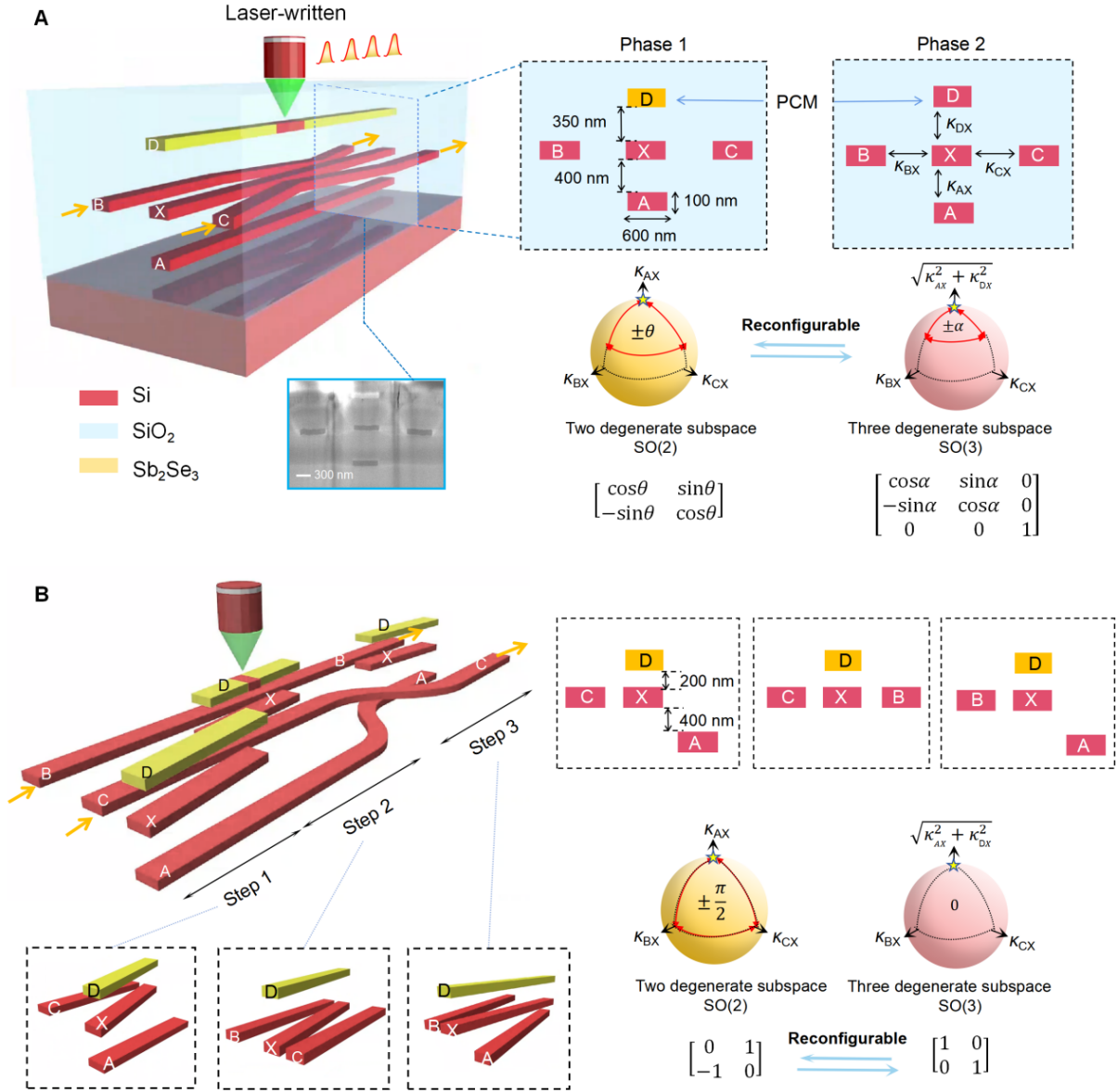

**Fig. 1. Schematic of reconfigurable non-Abelian holonomy. (A)** Structure of reconfigurable SO(2) holonomy in three-layer hybrid integrated photonics platform incorporating PCM. All waveguides share an identical cross-section (600 × 100 nm$^2$). The top waveguide is made of Sb$_2$Se$_3$, while the remaining waveguides are made of amorphous silicon. The crystalline and amorphous states of Sb$_2$Se$_3$ is tuned by hot plate and laser-direct-written, which modulate the number of degenerate spaces, further construct different SO(2) holonomy. **(B)** Reconfigurable two-mode braiding, which consists of three STIRAP processes, and the geometric phase can be modulated from π/2 to 0.

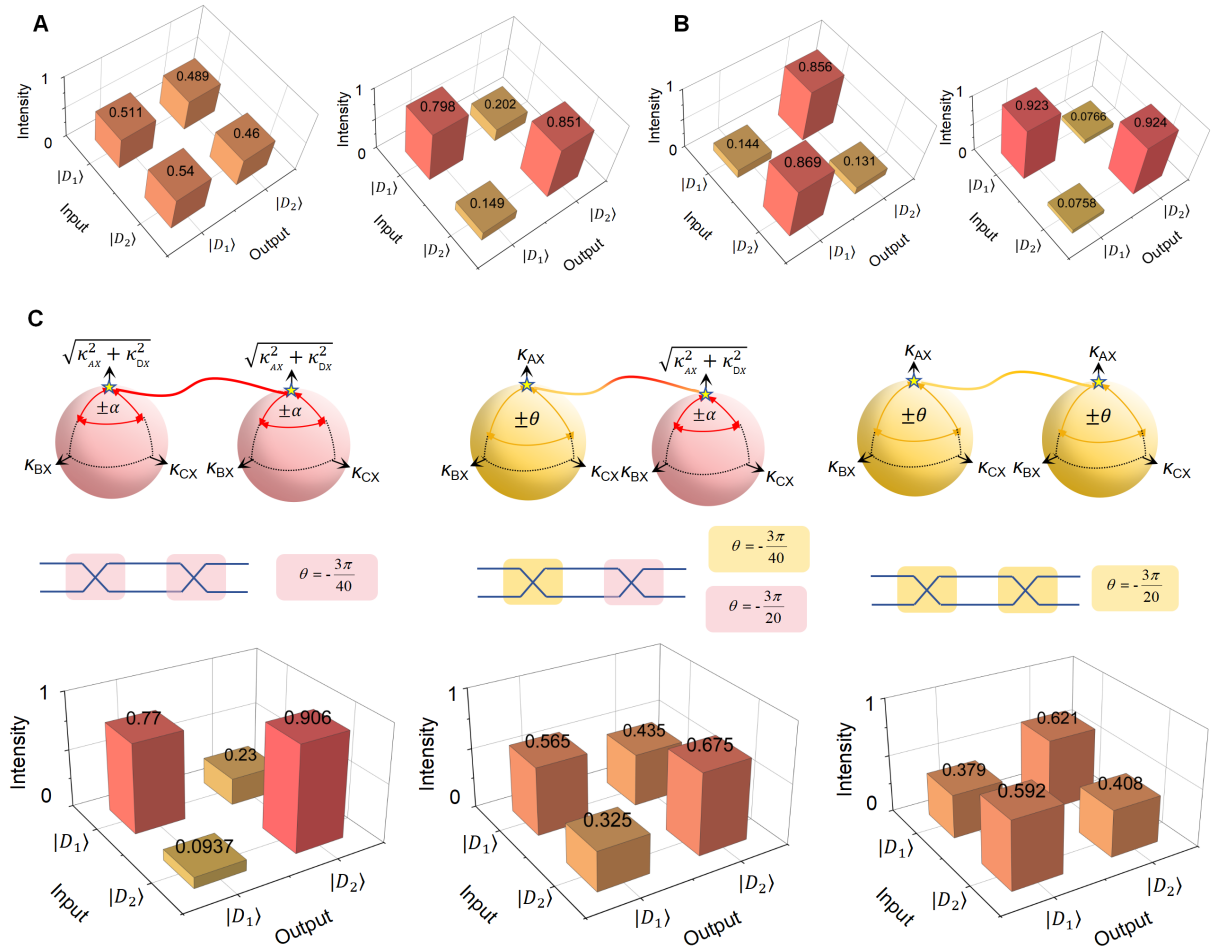

**Fig. 2. Experimental results of multi-level reconfigurable SO(2) holonomy.** (A) An ordinary two-level reconfigurable SO(2) holonomy. (B) Reconfigurable two-mode braiding. (C) An three-level reconfigurable SO(2) holonomy.

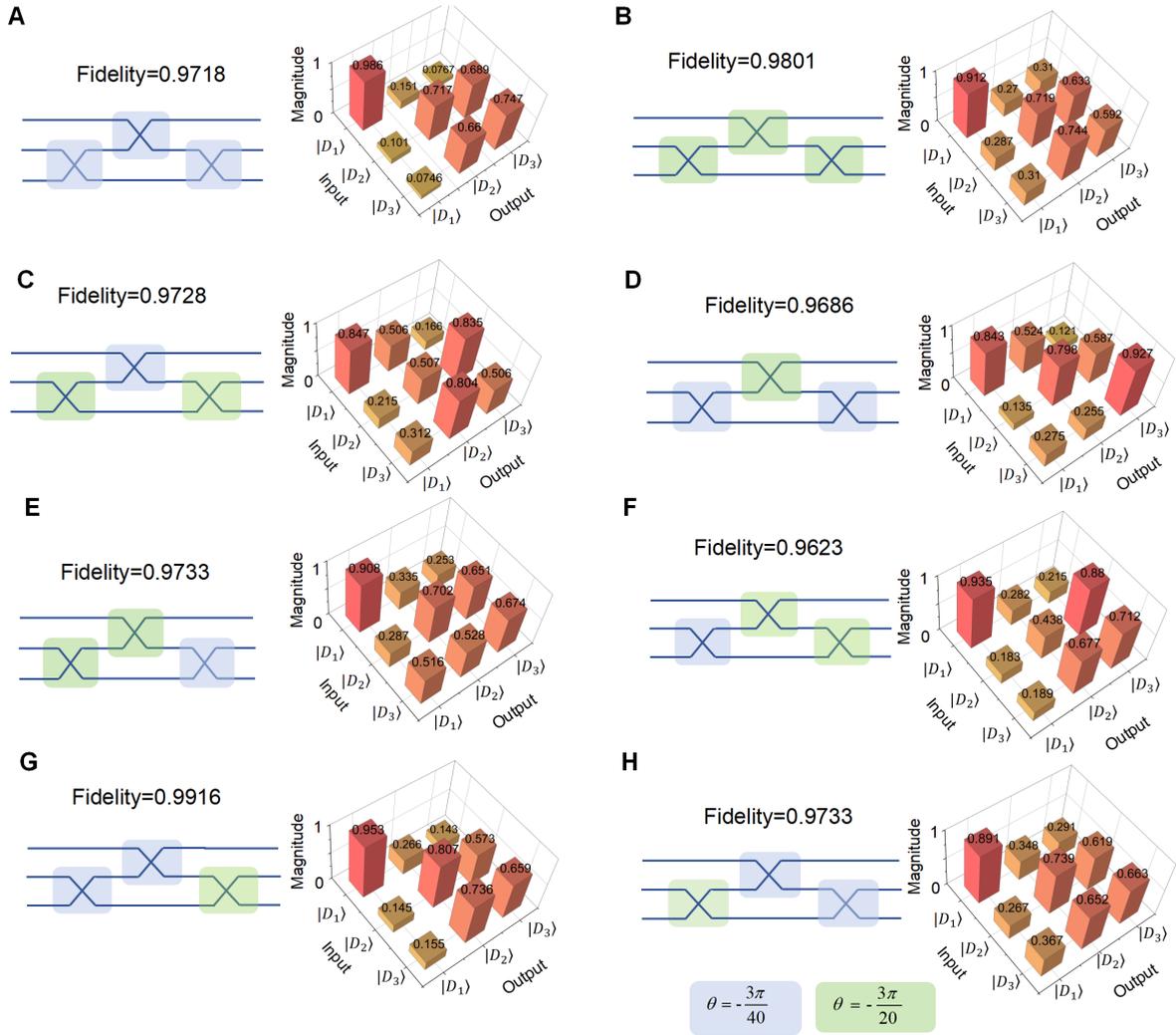

**Fig. 3. Three-bit (eight-level) reconfigurable SO(3) transformation synthesized using Givens rotations. (A-H)** The mathematical models, experimentally measured matrix elements magnitude, as well as the fidelity. The mathematical models illustrate different combinations of building blocks states.